\documentclass[preprint]{aastex}
\usepackage{emulateapj5}
\usepackage{apjfonts}
\usepackage{natbib}
\received{2001 May 1}
\revised{}
\accepted{}
\journalinfo{Astrophysical Journal Letters, submitted}
\submitted{Received 2001 May 1; Accepted May 22}

\begin{document}
\def\Ha{H$\alpha$ }
\def\Pa{Pa$\alpha$ }
\def\Hans{H$\alpha$}
\def\Pans{Pa$\alpha$}
\def\hst{{\it HST }} 
\def\gtorder{\mathrel{\raise.3ex\hbox{$>$}\mkern-14mu
             \lower0.6ex\hbox{$\sim$}}}
\def\ltorder{\mathrel{\raise.3ex\hbox{$<$}\mkern-14mu
             \lower0.6ex\hbox{$\sim$}}}
\def\asec{^{\prime\prime}}

\title{The Super Star Cluster NGC~1569-A Resolved on 
Sub-Parsec Scales \\
with {\it Hubble Space Telescope} 
Spectroscopy\footnotemark[1]}

\footnotetext[1]{Based on observations made with the {\it Hubble Space
Telescope}, which is operated by AURA, Inc., under NASA contract NAS5-26555.}

\author{Dan Maoz\footnotemark[2] \footnotemark[3], 
Luis C. Ho\footnotemark[4], and Amiel Sternberg\footnotemark[3]}

\footnotetext[2]{Department of Astronomy, Columbia University,
550 W. 120th St., New York, NY 10027; dani@wise.tau.ac.il}

\footnotetext[3]{School of Physics \& Astronomy, 
Tel-Aviv University, Tel-Aviv 69978, Israel;
amiel@wise.tau.ac.il}

\footnotetext[4]{The Observatories of the Carnegie Institution of Washington,
 813 Santa Barbara St., Pasadena, 
CA 91101; lho@ociw.edu}

\begin{abstract}
We present  3000--10000 \AA\ {\it HST}/STIS 
long-slit spectroscopy of the bright 
super star cluster A (SSC-A) in the dwarf starburst galaxy NGC~1569. 
 The 0\farcs05 \hst angular 
resolution allows, for the
first time, to probe for spatial variations in the stellar population
of a $\sim 10^6 M_{\odot}$ SSC. Integrated ground-based spectra of SSC-A have 
previously revealed young  Wolf-Rayet (WR) signatures that coexist 
with features from supposedly older, red supergiant (RSG), populations. We find
that the WR emission complexes come solely from the subcluster A2, identified
in previous \hst imaging, and are absent from the main cluster A1, thus 
 resolving the question of whether the WR and RSG features
    arise in a single or distinct clusters. 
The equivalent widths of the WR features in A2 ---
including  the {\ion{C}{4}}~$\lambda 5808$ complex which
we detect  in this object for the first time --- are
larger than previously observed in other WR galaxies.
Models with sub-solar metallicity, as inferred from the nebular emission
lines of this galaxy, predict much lower equivalent widths. 
On the ``clean'' side of A1, opposite to A2, we find no evidence
for radial gradients in the observed stellar population 
at 0\farcs05$<R<$0\farcs40 ($\sim
0.5$ to 5~pc), neither in broad-band, low-resolution,
spectra nor in medium-resolution spectra 
of the  infrared {\ion{Ca}{2}} triplet.
   
\end{abstract}


\section{Introduction}

A substantial fraction of the star formation in nearby starbursts,
be it in merging galaxies, dwarf galaxies, 
or early-type galaxies with circumnuclear rings, 
takes place in so-called super star clusters (SSCs; e.g., Whitmore et al.
1999; Hunter et al. 2000; Maoz et al. 2001, and references therein).
These are clusters of stars with total luminosities as high as
 $M_V =-15$ mag  ($L_V=1.3\times 10^8 L_{V \odot}$) and 
radii of order a few pc. SSCs present a 
mode of massive star formation distinct from the OB associations found in most
star-forming regions of quiescent galaxies. They are of interest for
understanding star formation at high redshift, which is dominated
by the starburst mode, and the formation of globular clusters, into which
SSCs may evolve if they contain sufficient numbers of low-mass stars, and
can survive disruption and evaporation. 

Two of the nearest ($2.5\pm0.5$~Mpc; O'Connell, Gallagher, \& Hunter 1994)
and best-studied SSCs are clusters A and B
 in the amorphous dwarf
galaxy NGC~1569 (Arp \& Sandage 1985; Melnick, Moles, \& Terlevich 1985). 
NGC~1569-A is one of only 
several SSCs that have a kinematic mass estimate, $\approx 10^6 M_{\odot}$
(Ho \& Filippenko 1996; Sternberg 1998). 
Prada, Greve, \& McKeith (1994)
detected the {\ion{Ca}{2}} $\lambda\lambda 8498, 8542, 8662$ absorption
triplet, which forms in the atmospheres of red giants and supergiants,
 in cluster A. 
Based on the equivalent width (EW) of these lines, they estimated an 
age $> 13$ Myr.
Gonz\'alez Delgado et al. (1997), however, detected in the SSC-A spectrum 
the bump around 4686~\AA\ generally ascribed to {\ion{He}{2}}, {\ion{C}{3}}, 
{\ion{C}{4}}, {\ion{N}{3}}, and {\ion{N}{5}} emission from Wolf-Rayet 
(WR) stars. WR stars
are the evolved stages
of stars more massive than $\sim 30 M_{\odot}$, and are thought to
exist only when a burst is $3- 5$~Myr old. To explain this,  
Gonz\'alez Delgado et al. (1997) proposed that there have been several 
succesive bursts of star formation in SSC-A. 
De Marchi et al. (1997) analyzed post-refurbishment
{\it Hubble Space Telescope (HST)}
 images and showed that SSC-A has two peaks in its light distribution,
separated by 0\farcs18, which they designated A1 and A2, and
postulated that these are two distinct clusters, with the
WR features coming from one cluster and the red supergiant signatures
from the other.
More recently, Hunter et al. (2000) obtained deeper \hst images in several
bands, and argued against the proposed solution of De Marchi et al. (1997),
based on the finding that A1 and A2 have similar colors. They noted that
WR stars and red supergiants are
expected to co-exist, to some degree, when a cluster is about 5~Myr old.
Nonetheless, Buckalew et al. (2000) have found evidence,
 based on \hst {\ion{He}{2}} narrow-band imaging, that the WR emission
is concentrated in A2. 

More generally, how SSCs form and survive as bound objects is an
unsolved problem, given the 
rapid ejection of gas from the proto-clusters by jets,
ionizing radiation, stellar winds, and subsequent supernovae.
Some clues may be found in the spatial structure of the nearest SSCs.
%
NGC~3603 in the Milky Way and 
R136 in the  Large Magellanic Cloud (LMC) 
show some evidence for segregation of the stellar
population, with the most massive stars concentrated at smaller radii
(Moffat, Drissen, \& Shara 1994;
Brandl et al. 1996; Hunter et al. 1996). 
 It is thought that
the clusters are too young for dynamical mass segregation to have occurred
(Bonnell \& Davies 1998).
The segregation therefore may reflect a gradient in age or in initial 
mass function (IMF) properties,
but could also be a statistical effect,
due to the rarity of high-mass stars in the outer, low stellar-density regions
of the clusters. Current
estimates for the masses of these clusters are $\sim 10^4 - 10^5 M_{\odot}$
(Drissen et al. 1995).
The spatial population profile of massive, $10^6 M_{\odot}$
SSCs such as NGC~1569-A  has never been observed. 
This Letter presents resolved \hst long-slit spectroscopy of NGC~1569-A to
address several of these issues. 

\section{Observations and Reduction}

We observed NGC~1569 on 28 November 2000 UT, using the Space Telescope Imaging
Spectrograph (STIS) on \hst. The $52''\times$ 0\farcs1 slit was centered
on SSC-A, and the spacecraft was oriented with the intention 
that clusters A1 and  A2, and cluster B,
 $7''$ to the south-east, would all
fall within the narrow slit. Unfortunately, a 0\fdg45 error in the header
orientation information of a WFPC2 image of the galaxy from 1996, combined
with a 0\fdg33 error in determination of the STIS slit orientation in
the focal plane at the time the observation was planned, resulted in a 
total orientation error of 0\fdg78.  The 0\farcs1-wide slit 
was misplaced 0\farcs1 southwest from the center of SSC-B, 
effectively missing it.
A repeat observation of SSC-B is planned, and in this Letter 
we will analyze only the spectra of SSC-A.

\vspace{0.1in}
\centerline{\includegraphics[angle=0,width=0.8\columnwidth]{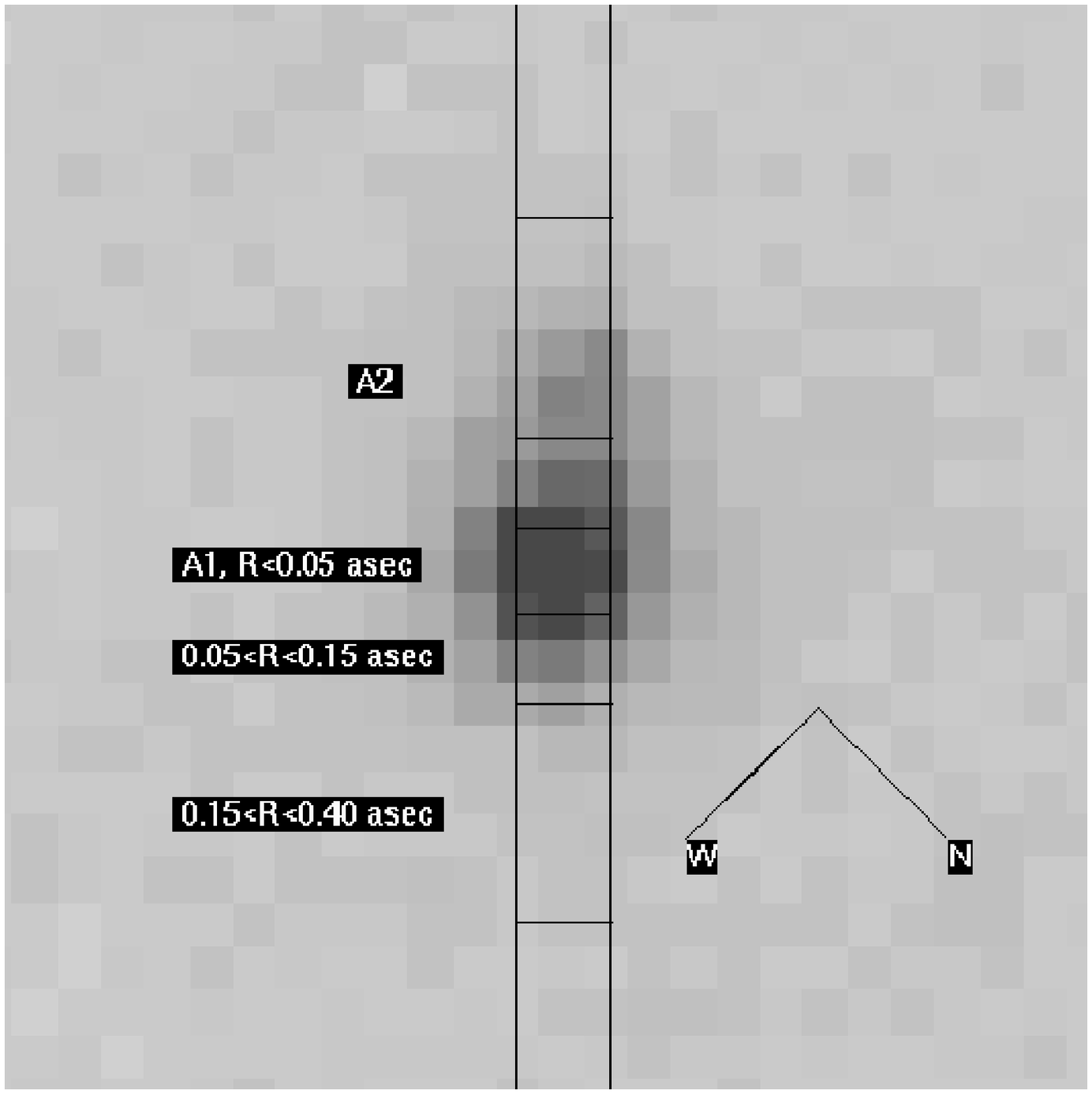}}
\begin{quote}
\baselineskip3pt
{\footnotesize Fig.~1-- 
Section (1\farcs15$\times$1\farcs15) of
STIS acquisition image of NGC~1569-A. The slit position is shown, and
the various spatial bins for which spectra were extracted are labeled.}
\end{quote}

Exposure times totaled 32.4 min in the G750L grating (5250--10250 \AA,
10 \AA\ resolution full width at half maximum [FWHM]), 
43.2 min in the G430L grating (2910--5714 \AA,
5.5 \AA\ resolution), and 93.6 min in the G750M grating (8276--8842 \AA, 
1.1 \AA\ resolution). 
The STIS plate scale is 0\farcs05 pixel$^{-1}$. 
Flat-field exposures were taken with the G750L and G750M
gratings for use in fringe correction. 
The data were initially processed by
the STScI pipeline, producing geometrically corrected, flux- and 
wavelength-calibrated two-dimensional spectral images at each of three
dither positions.
The fringe flats were normalized and then scaled to give
the best suppression of the fringing in the science images.
The images at the individual dither positions were registered
and combined. Spectra of each of the resolved spatial bins of SSC-A were
then obtained by summing individual rows. The five spatial bins chosen
were a central bin, consisting of the two rows with peak flux ($R<$0\farcs05),
two intermediate-radius bins, each consisting of two rows above and below
the central bin (0\farcs05$<R<$0\farcs15), and two outer bins
(0\farcs15$<R<$0\farcs40). The outer bin on the south-east side of SSC-A
is dominated by flux from SSC-A2, while the center and north-west bins are
dominated by flux from SSC-A1. 
Figure 1 shows the location of the rectangular
sections of SSC-A we have thus probed.

Despite the geometric correction applied to the data in the pipeline, some
residual curvature remains in the spectral images. 
Combined with the unavoidable
variations in point-spread function (PSF) along the large wavelength interval
 of the low-resolution spectra, this produces conspicuous
artificial spectral variations
between different rows. 
To correct for these effects, we obtained the G430L
and G750L archival data of two white dwarfs, GRW+70D5824 and Feige~110,
respectively,
which were observed with the same STIS setup. We simulated an A1+A2 cluster
pair having {\it no} spatial color gradient
 by convolving the spectral images of the stars
 with Gaussians, scaling, shifting,
and adding the images, according to the widths and relative fluxes measured
in WFPC2 images by Hunter et al. (2000). 
We verified that the cross-dispersion
profile at 5500~\AA\ of the simulated data matched closely 
that of the real data, 
which have FWHM of 0\farcs20$\pm$0\farcs02.
 We then 
extracted spectra in the same spatial bins as above, normalized
 by the total spectrum of the simulated cluster. The resulting
correction function was heavily smoothed to eliminate residual spectral 
features from the white dwarfs, and each spatial bin in the real data was
divided by its corresponding correction function. The correction functions
are insensitive, at a level of 
a few percent or less, to the details of the simulation,
 as long as it includes 2 sources with approximately the correct scaling
and the total profile has
0\farcs15$<$FWHM$<$0\farcs27.   

Background emission from the sky and from the galaxy in the neighborhood of the
cluster is negligible.
To obtain a ``clean''
spectrum of A2, the spectrum of the A1 bin diametrically opposed to
it was subtracted from the A2 bin. 
Residual cosmic rays and bad pixels
in the final spectra were identified by eye and corrected by interpolation
between adjacent pixels.
There is agreement in slope and flux level
between the G430L and G750L spectra  in the overlap region 
at 5250~\AA~ to 5714~\AA~
to the few-percent level,
except for the central bin spectra, where the G430L spectrum must be scaled
up by 15\% to match the G750L data. After accounting for the 40\% 
transmission of the total  light by the 0\farcs1 slit, the 
total SSC-A flux matches well the measurements by Gonz\'alez Delgado et al. (1997) 
and De Marchi et al. (1997). 
  
\section{Analysis and Discussion}

\subsection{Subcluster A2 and the Wolf-Rayet Emission}


Figure 2 shows the final low-resolution spectra for each of the
spatial bins. 
To allow easy comparison of the spectral shapes,
the spectra have been multiplicatively scaled up to match the central-bin
spectrum in the blue region, and then shifted up or down by an additive 
constant. The spectrum of the A2 bin is shown before subtraction of
the northwest, diametrically-opposed, bin of A1 (NW 0\farcs15$<R<$0\farcs40).
The EWs of the main emission and absorption features are
listed in Table 1.

Gonz\'alez Delgado et al. (1997)
 first reported the WR emission bump at 4686 \AA\
in the integrated spectrum of SSC-A.
It is clear from Figure~2 that this feature, as well as the strong
{\ion{C}{4}}~$\lambda 5808$ \AA\ WR complex which we identify for the first time
in this object, 
come almost exclusively from SSC-A2. Indeed, the weakness or absence
 of the WR features in 
the A1 bins demonstrates the angular resolving power of this observation.
These results resolve the issue of the co-existence, in SSC-A,
 of WR with red supergiant signatures, confirming the
interpretation proposed by De Marchi et al. (1997) and the
narrow-band imaging results of Buckalew et al. (2000). It is unknown whether,
at the 0\farcs18 (2.2 pc) separation,
A2 is a separate cluster seen in projection, or rather a
 physical neighbor or substructure to A1. In either case, the data show
that A2 constitutes a distinct stellar population. 
Note that, apart from the WR features in A2,
the stronger Balmer-line emission in A2, and a possible difference in
the spectral shape at $\lambda<3600$~\AA, the spectra of A1 and A2 are 
 quite similar. This confirms and explains the observation
by Hunter et al. (2000) of similar optical colors in A1 and A2.

\centerline{
\includegraphics[angle=0,width=4.5in]{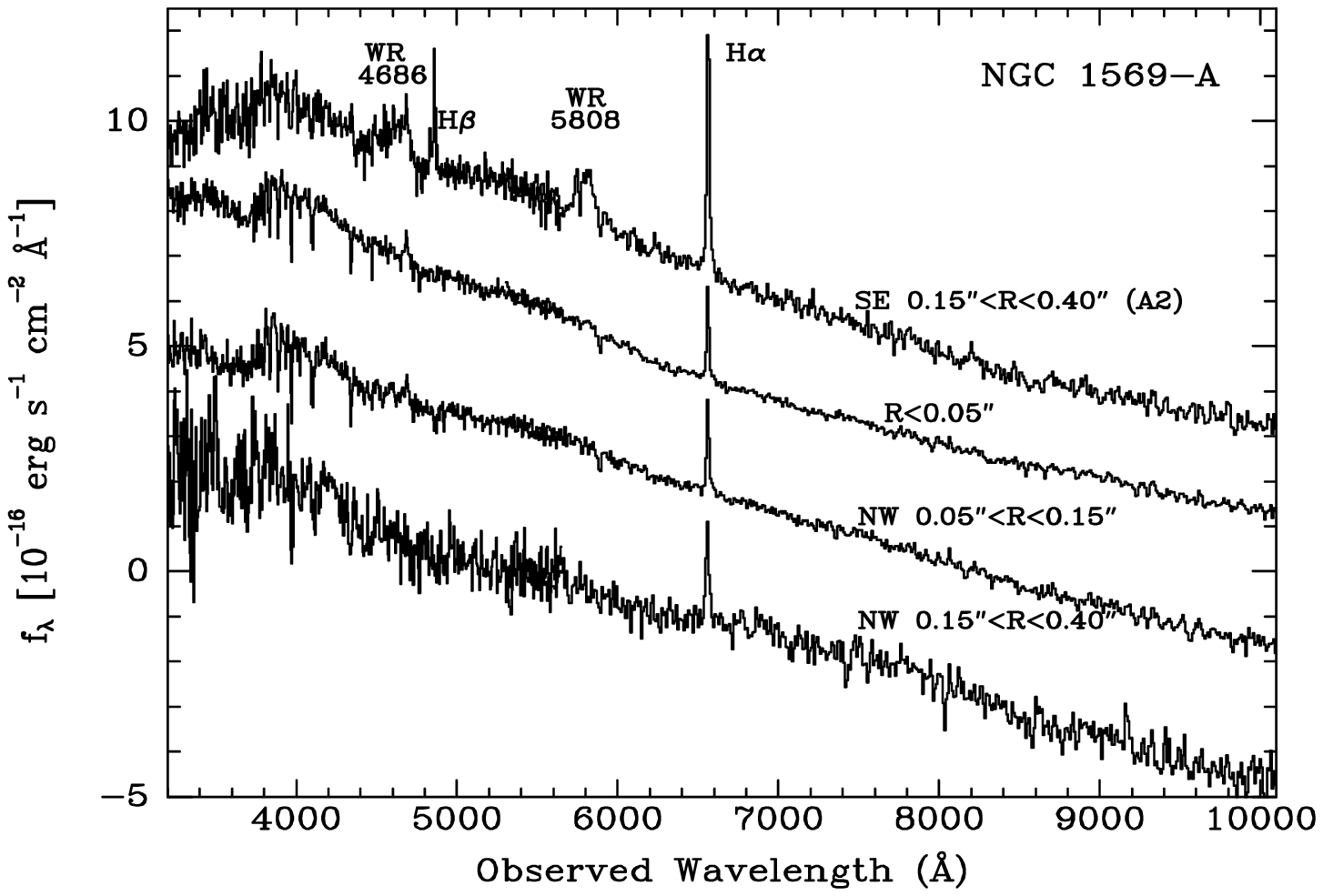}}
\begin{quote}
{\footnotesize Fig.~2-- 
Broad-band spectra for each of the
spatial bins of NGC~1569-A. The flux scale corresponds
to the second spectrum from the top (the central, $R<$0\farcs05, bin of A1).
Note the strong
WR emission that comes almost exclusively from the A2 bin, and the similarity
of the continuum shape among all bins. 
Scaling and shifting that has been
applied to the spectra, from top to bottom, is
 $2.5 f_{\lambda}+2.2\times 10^{-16}$;
$f_{\lambda}+0$; $2.25 f_{\lambda}-3.5\times 10^{-16}$; 
and $5.6 f_{\lambda}-6.5\times 10^{-16}$.}
\end{quote}

The flux in the 4686 \AA\ feature is consistent with the measurement
by Gonz\'alez Delgado et al. (1997), who concluded it requires the 
luminosity of 20--40 WNL stars. However, now that this emission is
localized to the relatively faint sub-clump A2, the EW
of the WR features becomes huge. After subtraction of the 
background from A1, 
we measure EW(4686)=33$\pm9$~\AA~ 
and EW(5808)=32$\pm9$~\AA. For comparison, Schaerer, Contini, \& Kunth (1999)
have measured in five WR galaxies typical values of 
EW(4686)=4--10~\AA\ 
and EW(5808)=2--6~\AA. They find the highest values, with
EW(5808)$\sim 12$~\AA, in the galaxy Tol~89.
Schaerer \& Vacca (1998) have constructed 
evolutionary synthesis models for instantaneous starbursts and predicted the
strengths of the WR features as a function of starburst age and metallicity.
As seen, for example, in Figure~11 of
 Schaerer \& Vacca (1998), the WR emission peaks briefly between
ages of 3--5 Myr, and its strength is proportional to the metallicity $Z$.
For $Z=0.2 Z_{\odot}$,  they predict that the EW of either
of the WR bumps never rises above 8 \AA, and for $Z=0.4 Z_{\odot}$, it 
has a maximum of 11 \AA. Only at $Z=2Z_{\odot}$ do the EWs ever reach 
EW(4686)=30~\AA\ and EW(5808)=25~\AA.

However, several measurements of the gas chemical abundances in NGC~1569
have yielded a significantly
sub-solar metallicity, of $Z\approx 0.25 Z_{\odot}$ (Calzetti, Kinney, 
\& Storchi-Bergmann 1994; 
Kobulnicky \& Skillman 1997; Gonz\'alez Delgado et al.
1997). The large EWs of the WR features we find in A2 would therefore 
indicate that,
contrary to the line-emitting gas from which they were recently formed, 
the stars in A2 have solar
or supersolar abundances. Alternatively, there may be 
a problem with the models,
in that the WR lifetimes depend on uncertain mass-loss rates and
    the effects of stellar rotation (Maeder \& Meynet 2000).
Note that, in a non-instantaneous starburst,
 the EW of the WR features
will necessarily be smaller, thus exacerbating the problem. 
If A2 is not a separate cluster, or a subcluster of A, but rather
a particular physical region of cluster A, then it is not clear that
subtraction of the A1 background is justified. In that case, the EWs (and the 
errors)
of the WR features are halved, but are still double the values expected
for the metallicity of the gas.
 
\subsection{Cluster A1 Population and Radial Color Gradient}

The data for the ``clean,'' northwest, side of SSC-A1  in Fig. 1 reveal the 
spectrum of A1, uncontaminated by A2. We defer a detailed analysis
and comparison with models to a future paper. However, except for the
WR features which we have shown come from A2, the main characteristics
of the spectrum are unchanged in comparison to the integrated spectra
analyzed by previous ground-based studies 
(e.g., Gonz\'alez Delgado et al. 1997).
Specifically, the blue slope and the weakness of the Balmer jump
still point to an age of $\sim 5$ Myr, and the presence of the {\ion{Ca}{2}} triplet 
absorption rules out a younger age. 

It is remarkable that A1 and A2 appear to have similar ages of $\sim 5$~Myr,
 but that one has WR emission and the other does not. A possible explanation
is that A1 has an IMF with an upper cutoff 
$\ltorder 30 M_{\odot}$, and therefore does not form WR stars. 
A higher mass cutoff in A2
could be related to its anomalously high metallicity, suggested
by the large EWs of its WR features.  
 Alternatively, A1 may also have high, approximately solar, metallicity,
but is slightly older than A2, say $\sim 7$~Myr, by which time it would
have exited its WR phase (Schaerer \& Vacca 1998).
Finally, A1 may have the normal, sub-solar metallicity of the galaxy,
in which case it would just be exiting its WR phase already at 5~Myr. 
Thus A1 and A2 must have widely different IMFs, or widely different abundances,
or similar, anomalously high, abundances but slightly different ages.
  
The northwest A1 data also allow us to examine, for the first time, the radial 
gradient in the stellar population of an individual high-mass SSC.
As seen in Figure~2, there is remarkable similarity in the spectra of A1
in the three radial bins shown. This indicates that, at least on its northwest
side, A1 is extremely homogeneous. Considering the small age, the
formation of stars in A1 must have been highly synchronized throughout 
its volume. In this respect, SSC-A1
appears to differ from other, lower mass,  SSCs that 
possibly do show population segregation (see \S1).

\subsection{Calcium Triplet Spectra}

Figure 3 shows a region of the medium-resolution G750M spectra centered
on the near-infrared {\ion{Ca}{2}}
triplet.
The spectra of the outer bins are too noisy to reveal unambiguously 
the {\ion{Ca}{2}} triplet, and are not shown. 
Figure~3 and Table 1 show
 that there is no significant difference in {\ion{Ca}{2}} triplet absorption
between the central and intermediate radial bins in SSC-A1.
 This result confirms the lack of a population gradient
seen in the low-resolution, broad-band spectra.     
Although the strength of the {\ion{Ca}{2}} triplet
absorption, which comes first from red supergiants and at a later stage
from red giants,
has been used to argue for a large age for SSC-A (Prada et al. 1994), the data
are also consistent with an age of $\sim 5$~Myr.
Garc\'ia-Vargas, Mollan, \& Bressan (1998) have used evolutionary synthesis
models to predict  the EW for the sum of the two strongest triplet lines,
at 8542~\AA~ and  8662~\AA. The summed EW value seen
in NGC~1569, of $\sim$3--5 \AA, 
is generally lower than the value predicted for starbursts 
older than 6~Myr, regardless of metallicity (cf. B\"oker et al. 2001 for
caveats on the accuracy of the low-metallicity models).  The observed 
EW, on the other hand, briefly matches the model predictions for cluster ages 
5--6 Myr, when the red supergiants first form.

\hspace*{-0.4in}
\includegraphics[angle=0,width=3.5in]{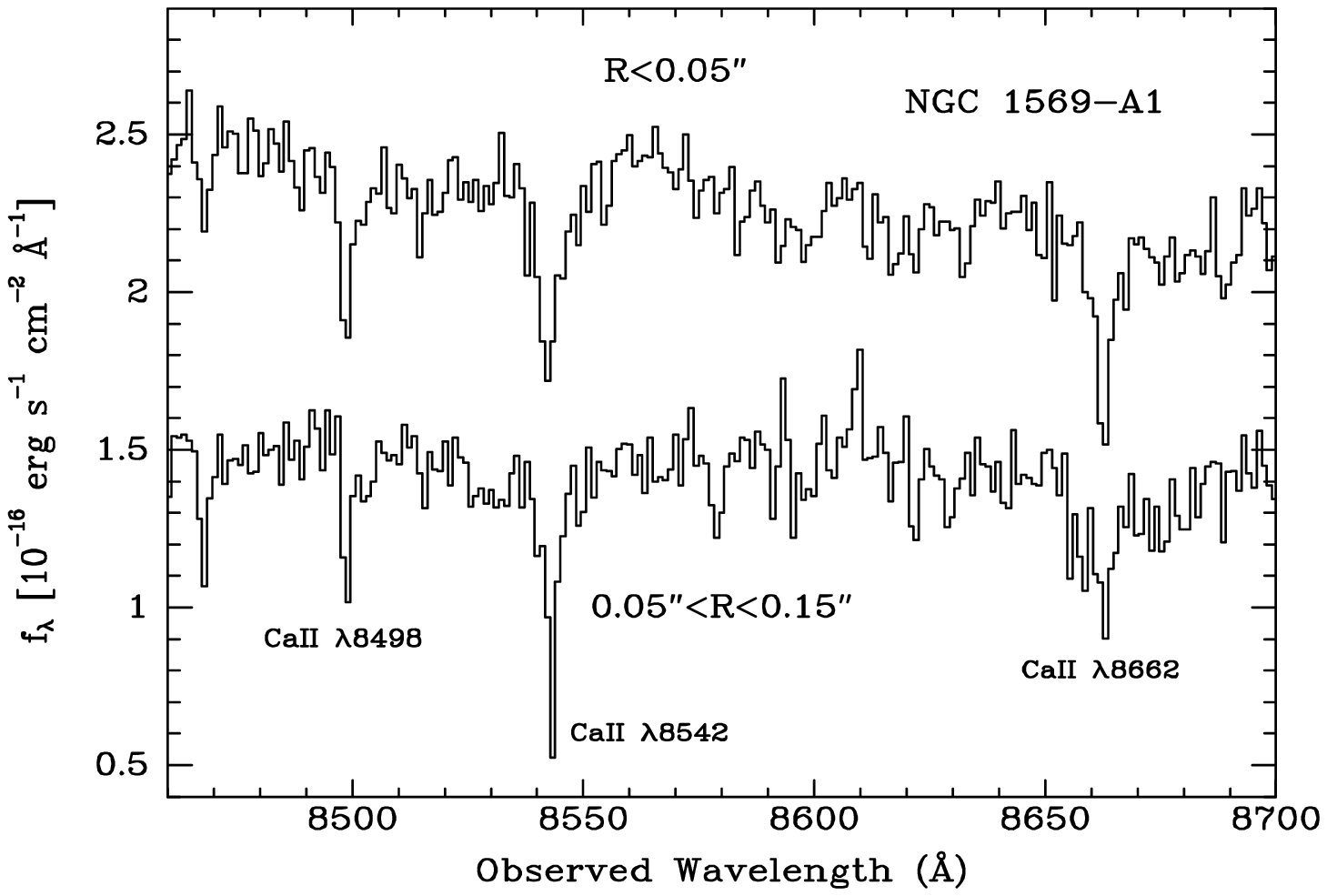}
\begin{quote}
\baselineskip3pt
{\footnotesize Fig.~3-- Medium-resolution spectra centered
on the near-infrared {\ion{Ca}{2}} triplet absorption.
The lower spectrum, which is the sum of the spectra of the two bins
straddling the central bin, has been shifted down by 0.8 units.
Note the similarity of {\ion{Ca}{2}} triplet absorption strength at the different radii.}
\end{quote}
\smallskip
\bigskip
\bigskip

\subsection{Summary}

We have obtained spectra of NGC~1569-A that are resolved
at the sub-pc scale, the 
first time a $10^6 M_{\odot}$ SSC is observed this way.
The data clearly show that the WR emission from this SSC is confined
to subcluster A2. The EWs of the WR features are much larger than
previously observed in other WR galaxies, and are several times larger
than the maximum values predicted from models with sub-solar metallicities,
as measured
in the gas of this galaxy. This problem is particularly severe
 if A2 is a separate
structure (as opposed to a region inside A1), justifying the subtraction
of the underlying light of A1 from A2.  The continuum slope, the Balmer jump,
the {\ion{Ca}{2}} absorption and WR emission EWs, are all consistent with an age
of $\sim 5$~Myr for the bursts that formed both A1 and A2.  While there is a 
sharp difference in WR population between A1 and A2, 
there is no significant spatial 
gradient in the stellar population on the NW side of A1.

We conclude that, either A2 has an anomalous, super-solar metal abundance,
or the models have seriously underpredicted the WR emission strength for
sub-solar metallicities. The weakness of WR features in A1, despite the similar
age to A2, could be caused by a much lower IMF upper-mass cutoff in A1,
a much lower metallicity in A1, or a slightly larger age for A1. Finally,
the homogeneity of A1 is distinct from the possible radial mass segregation 
reported in some lower mass SSCs in the Galaxy and the LMC.
Further analysis is required to determine the degree to which the apparent
uniformity may be set by the most massive stars, which dominate the light.  
If real, it
needs to be reproduced by models of SSC formation and evolution.

\acknowledgements
This work was supported by grant GO-8293 from the Space Telescope
Science Institute, which is operated by AURA, Inc., under NASA contract NAS
5-26555. 

\end{multicols}
\begin{deluxetable}{rrrrr}
\tiny
\tablewidth{0pt}
\tablecaption{NGC~1569-A: Emission and Absorption Equivalent Widths}
\tablehead{\colhead{}&\colhead {A2} & \multicolumn{3}{c}{A1} \\
\colhead {}&\colhead {}&
\colhead {$R<$0\farcs05}&\colhead {0\farcs05$<R<$0\farcs15}
&\colhead {0\farcs15$<R<$0\farcs40}\\}
\startdata
H$\delta  \lambda 4102$     &$-0.5\pm0.5$ &$-2.8\pm0.4$ &  $-3.0\pm0.7$  & $-1\pm1$ \\
H$\gamma  \lambda 4340$       &$1.2\pm0.8$&$-2.0\pm0.3$ &  $-2.5\pm0.5$  & $-2\pm1$ \\
WR       $\lambda 4686$       & $33\pm9$  &$ 1.9\pm0.5$ &  $ 1.1\pm0.5$  & $<1.5  $ \\
H$\beta   \lambda 4861$       & $11\pm1$  &$-0.3\pm0.1$ &  $-0.9\pm0.3$  & $0\pm1 $ \\
WR       $\lambda 5808$       & $32\pm9$  &$<3        $ &  $<3        $  & $<8    $ \\
H$\alpha  \lambda 6563$       & $56\pm6$  &$8.8\pm0.5 $ &  $8.5\pm1.0 $  & $10\pm2$ \\
Ca II    $\lambda 8498$       &\nodata    &$-1.1\pm0.2$ &  $-0.9\pm0.2$  & \nodata  \\
Ca II    $\lambda 8542$       &\nodata    &$-2.4\pm0.6$ &  $-1.5\pm0.5$  & \nodata  \\
Ca II    $\lambda 8662$       &\nodata    &$-1.7\pm0.5$ &  $-1.7\pm0.6$  & \nodata  \\
\enddata
\tablecomments{Equivalent widths in \AA. Positive values are for emission and negative
for absorption. WR denotes the broad Wolf-Rayet emission complexes roughly centered on the wavelengths listed. Errors correspond to $\sim 2\sigma$ and are dominated by continuum placement
uncertainty.}
\end{deluxetable}						     

\begin{multicols}{2}


\begin{references}

\reference{}{Arp, H., \& Sandage, A. 1985, AJ, 90, 1163}

\reference{} 
B\"oker, T., van der Marel, R.~P., Mazzuca, L., Rix, H.-W.,
Rudnick, G., Ho, L.~C., \&  Shields, J.~C. 2001, AJ, 121, 1473

\reference{}
Bonnell, I.A., \& Davies, M.B. 1998, MNRAS, 295, 691

\reference{}{Brandl, B., et al. 1996, ApJ, 466, 254}

\reference{}Buckalew, B.\ A., Dufour, R.\ J., Shopbell, P.\ L., \& Walter, D.\ K.\ 
2000, \aj, 120, 2402 
 
\reference{}{Calzetti, D., Kinney, A.\ L.\ \& Storchi-Bergmann, T.\ 1994, \apj, 429, 582}


\reference{}
De Marchi, G., 
Clampin, M., Greggio, L., Leitherer, C., Nota, A., \& Tosi, M.\ 1997, 
\apjl, 479, L27 

\reference{} 
Drissen, L., Moffat, A.\ F.\ J., Walborn, N.\ R., \& Shara, M.\ M.\ 1995, 
\aj, 110, 2235 






\reference{} Garc\'ia-Vargas, M.\ L., Molla, M., \& Bressan, A.\ 1998, \aaps, 130, 513 


\reference{}
Gonz\'alez Delgado, R.\ M., Leitherer, 
C., Heckman, T., \& Cervi\~{n}o, M.\ 1997, \apj, 483, 705 

\reference{}{Ho, L.~C., \& Filippenko, A.~V. 1996, ApJ, 466, L83}


\reference{}{Hunter, D. A., O'Connell, R. W., \& Gallagher, J. S. 1994, AJ, 108, 84}

\reference{}Hunter, D.\ A., O'Connell, R.\ W.,
Gallagher, J.\ S.\ \& Smecker-Hane, T.\ A.\ 2000, \aj, 120, 2383

\reference{} 
Hunter, D.\ A., O'Neil, 
E.\ J., Lynds, R., Shaya, E.\ J., Groth, E.\ J., \& Holtzman, J.\ A.\ 1996, 
\apjl, 459, L27 

\reference{}Kobulnicky, H.\ 
A.\ \& Skillman, E.\ D.\ 1997, \apj, 489, 636 


\reference{}Maeder, A. \& Meynet G. 2000,
    ARAA 38, 143

\reference{}{Maoz, D., Barth, A.~J., Ho, L.~C., Sternberg, A., 
\& Filippenko, A.~V. 2001, AJ, in press ((astrop-ph/0103213) }


\reference{}{Melnick, J., Moles, M., \& Terlevich, R. 1985, A\&A, 149, L24}

\reference{} 
Moffat, A.\ 
F.\ J., Drissen, L., \& Shara, M.\ M.\ 1994, \apj, 436, 183 

\reference{}{O'Connell, R. W., Gallagher, J. S., \& Hunter, D. A. 1994, ApJ, 433, 65}


\reference{} 
Prada, F., 
Greve, A., \& McKeith, C.\ D.\ 1994, \aap, 288, 396 


\reference{} Schaerer, 
D., Contini, T., \& Kunth, D.\ 1999, \aap, 341, 399 

\reference{} Schaerer, D.\ \& 
Vacca, W.\ D.\ 1998, \apj, 497, 618 



\reference{} Sternberg, A. 1998, ApJ, 506, 721

\reference{}{Whitmore, B. C., Zhang, Q., Leitherer, C., Fall, S.~M., Schweizer, F. 
\& Miller, B.~W. 1999, AJ, 118, 1551}
\end{references}
\end{document}